\documentclass[prb,twocolumn,amsmath,amssymb,superscriptaddress,citeautoscript,longbibliography]{revtex4-2}
\usepackage[utf8]{inputenc}
\usepackage[unicode=true,bookmarks=true,bookmarksnumbered=false,bookmarksopen=false,breaklinks=false,pdfborder={0 0 0},pdfborderstyle={},backref=false,colorlinks=true,]{hyperref}
\hypersetup{linkcolor=blue,citecolor=blue,urlcolor=blue}
\usepackage{lipsum}%
\usepackage{amsmath}%
\usepackage{amsfonts}
\usepackage{amssymb}
\usepackage{mathtools}
\usepackage{graphicx}
\usepackage{stackrel}
\usepackage[]{color}
\usepackage[left=1.50cm, right=1.00cm, top=2.00cm, bottom=2.00cm]{geometry}
\usepackage[]{bm}
\usepackage{siunitx}

\DeclareMathOperator{\sign}{sgn}
\DeclareMathOperator{\hc}{H.c.}
\DeclareMathOperator{\Res}{Res}
\graphicspath{{./figs/}}
\allowdisplaybreaks[1]

\begin{document}

\title{Bosonization solution of the Kondo lattice in a Luttinger liquid}

\author{Tom{\'a}s Bortolin}
\affiliation{Departamento de F\'{\i}sica, Universidad Nacional de La Plata, cc 67, 1900 La Plata, Argentina.}

\author{C. J. Bolech}
\affiliation{Department of Physics, University of Cincinnati, Cincinnati, Ohio 45221, USA}

\author{Nayana Shah}
\affiliation{Department of Physics, Washington University in St.\,Louis, St.\,Louis, Missouri 63130, USA}

\author{An\'{\i}bal Iucci}
\affiliation{Instituto de F\'{\i}sica de La Plata - CONICET, Diag 113 y 64, 1900, La Plata, Argentina.}
\affiliation{Departamento de F\'{\i}sica, Universidad Nacional de La Plata, cc 67, 1900 La Plata, Argentina.}

\begin{abstract}
%\textcolor{cyan}{\lipsum[1-1]}
We address the physics of a regular arrangement of independent magnetic impurities embedded in a band of interacting electrons. We focus on the one-dimensional case that can be studied using bosonization and in which the electron bulk is described by a Luttinger liquid. The impurity spins interact with the electrons via magnetic exchange that introduces the possibility of Kondo and Ruderman-Kittel-Kasuya-Yosida (RKKY) physics. We find that for two special values of the interactions, the model can be refermionized as a noninteracting electron band hybridized with a regular array of resonant levels. These solvable limits provide access to impurity correlators that correspond to either extended algebraic order or local screening. A physical picture emerges of how the interelectron interactions can stabilize either Kondo or RKKY physics depending on the sign of the interaction.
\end{abstract}

\date{\today}

\maketitle

\section{Introduction and Motivation}

The microscopic physics of solid-state materials is rich and varied. Thus classification plays a central role in sorting out phenomena and guiding our understanding. This effort goes hand in hand with the theories and models used to describe the different phenomenology. Focusing on electronic properties, one of the most basic divisions is Fermi versus non-Fermi liquids \cite{Pines}. The former are well described by models of weakly correlated electron-like quasiparticles, can be described from first principles using band theory and density-functional methods, and, despite the conceptual relative simplicity, they include interesting and rich phenomena like, for example, topological phases \cite{Vanderbilt}. Non-Fermi liquids, on the other hand, are a much more diverse group and, as it would be expected, not a single model or theory comprises the whole range of behaviors. By and large, these are strongly correlated electronic phases displaying a panoply of properties like long-range magnetic orders, interaction-driven insulators, unusual conductors, and even superconductors, to name a few. In this work, we shall turn our attention to low-dimensional non-Fermi liquids that host magnetic Kondo impurities.

\subsection{Contextual review}

Perhaps the simplest of the minimal models for this type of system, which nevertheless is widely expected to capture many of the rich phases just mentioned, is the Hubbard model and its generalizations \cite{hubbard1963,*hubbard1964,*hubbard1964a,*hubbard1965,*hubbard1967,*hubbard1967a,Mahan,Essler}
%%%
\footnote{The interest in the Hubbard model and in strongly correlated electrons in general became a central preoccupation of condensed-matter physics ever since the surprising discovery of the high-temperature cuprate superconductors in the mid 1980s \cite{bednorz1986}. 
In one dimension, the model is integrable and we have access to many of its properties (notably its thermodynamics) via the solution found by Lieb and Wu \cite{lieb1968}. This is nicely complemented by the bosonized description of the continuum limit of the model, the so-called g-ology model \cite{Giamarchi}. Such a description provides a global framework for the classification of electronic phases in one dimension; it shows that Fermi liquids are non-generic in this context (to be expected due to the severe restrictions in phase space for electron-electron scattering) and gives rise to the concept of \textit{Luttinger liquids} \cite{haldane1981}. The relevance of it for two dimensions, and cuprates in particular, has been greatly debated, starting with the ideas of the late P.\,W.\,Anderson \cite{anderson1990} and more recently in combination with the idea of electronic stripes \cite{kivelson1998,emery2000}.}.
%%%
While the Hubbard model focuses on capturing the effect of (short-range) electron-electron interactions, there are other routes to non-Fermi liquid behavior. In the context of a wide class of materials known as heavy fermions \cite{stewart1984,coleman1991}, the focus is on the interplay between localized and extended electronic degrees of freedom (or $d$ and $f$ electrons versus $s$ and $p$ electrons, respectively, in the chemist's terminology). The minimal model that captures such an interplay is an extension of the single-impurity Kondo model to regular arrays of magnetic impurities known as a \textit{Kondo lattice} \cite{tsunetsugu1997}.

\begin{figure}[th]
\begin{center}
\includegraphics{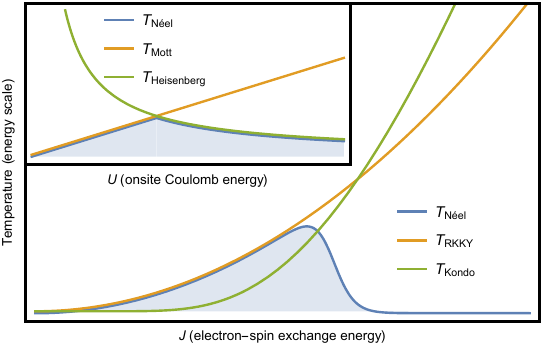}
\end{center}
\caption{\label{Fig:PhaseDiag}
Doniach's phase diagram. The crossing of the characteristic scales for Kondo screening [$\approx\sqrt{g}\exp(-1/g)$, where $g$ is the Kondo coupling $J$ multiplied by the electron density of states at the Fermi surface] and RKKY interactions [$\approx g^2/4$] indicates a competition between the tendency for the local moments to form singlets with the itinerant electrons or interact among themselves (mediated by the same itinerant electrons) to form a collective singlet. The latter corresponds to a long-range magnetically ordered state shaded under a N\'eel dome. Inset: schematic depiction of the ordering diagram in a 3D Hubbard model (depicted at the mean field level; for a more accurate picture see Ref.~\onlinecite{staudt2000}). The two ordering tendencies complement each other resulting in a dome region below both scales where one finds a long-range antiferromagnetic insulating phase.}
\end{figure}

The low-temperature physics of a Kondo-lattice system is determined by the competition between two distinct %effects
energy scales, a picture advanced early on by Doniach \cite{doniach1977,doniach1977a,jullien1977a,*jullien1977b,sullow1999,bernhard2000,*bernhard2011}
%%%
\footnote{Doniach pointed out that there are two emergent energy scales in the problem: one being the Kondo temperature, which is a decreasing exponential function of the reciprocal of the exchange coupling ($J$) between the localized moments and the itinerant electrons; the other the RKKY scale associated with the itinerant-electron-mediated ordering of the localized moments, which scales with $J^2$.}. 
%%%
For a relevant range of values of parameters (such as the density of electronic states at the Fermi surface), there exists a critical value of the magnetic exchange (say, $J_c$) at which the two energy scales cross each other. For $J<J_c$ the Ruderman-Kittel-Kasuya-Yosida (RKKY) scale is larger than the Kondo scale and, as the temperature is lowered, the local moments order magnetically before the Kondo effect would screen them. On the other hand, for $J>J_c$ the Kondo scale is the larger one and the moments are screened before they start to order. This competition is illustrated in the schematic diagram of Fig.~\ref{Fig:PhaseDiag} and can already be observed and rigorously studied in the case of the two-impurity Kondo model \cite{jayaprakash1981,jones1987,*jones1989,jones1988,affleck1995,gan1995a,*gan1995b,hallberg1997}
%%%
\footnote{These arguments do not require the local exchange interaction to be SU(2) symmetric, as observed experimentally, for instance, by neutron scattering in Ce-based heavy fermion materials \cite{goremychkin2000} and expected theoretically on account of the strong spin-orbit effect due to the $f$ electrons.}
%%%
\footnote{Meanwhile, in the half-filled Hubbard model, the interplay between Mott repulsion and N\'eel ordering due to direct magnetic exchange gives rise to a low-temperature phase characterized as an antiferromagnetic insulator (see the inset of Fig.~\ref{Fig:PhaseDiag}). In the laboratory, this corresponds to the state of the so-called \textit{parent compounds} of the cuprate superconductors --stoichiometric insulating materials that upon (oxygen) doping develop $d$-wave superconducting phases. Pairing is thought to take place as the effective 2D Hubbard models for the CuO$_2$ planes present in these compounds are driven away from half filling with electron or hole doping \cite{zhang1988}. The microscopic understanding for this basic phenomenology of cuprate superconductors is still incomplete and the subject of continuing studies.}.
%%%

The study of strongly correlated superconducting materials has all along been rich in interesting developments and relevant findings continue to happen every few years. After two decades of experimental search fueled by competing theoretical predictions \cite{anisimov1999,lee2004}, a group managed to synthesize thin films of \textit{nickelate superconductors} \cite{li2019,zeng2020}
%%%
\footnote{These are isostructural compounds to the high-temperature cuprate superconductors in which copper is replaced by its anteceding neighbor in the Periodic Table, nickel, and the crystal structure contains NiO$_2$ planes. A typical formula for the parent compound of these new superconductors is $R$NiO$_2$, where $R$ is a rare-earth element like Nd. Initially, the superconducting transition temperatures upon doping (with strontium, for example) were modest and increased somewhat when doped with praseodymium \cite{Osada2020}, but more recently there are reports of onset superconductivity above the boiling point of nitrogen in polycrystalline samples of La$_3$Ni$_2$O$_7$ under pressure \cite{Sun2023} (and more robust for oxygen-deficient samples \cite{Wang2024}).}. 
%%%
These discoveries are important because they allow for comparisons with the \textit{cuprate superconductors} that can help us sort out many issues that are still unresolved, like the role of magnetism in the genesis of superconductivity. Indeed, the parent nickelates are nonmagnetic and transport measurements show an upturn of the resistivity with lowering temperatures that is reminiscent of the Kondo effect (although other explanations are not yet ruled out) \cite{zeng2020}.

Leaving aside material-specific details, when comparing generic $R$CuO$_2$ versus $R$NiO$_2$ parent compounds the difference is that while in the former the relevant degrees of freedom are widely thought to be the Cu $3d$ and O $2p$ orbitals, in the latter there is debate between two possibilities: Ni $3d$ and O $2p$ on one hand or Ni $3d$ and $R$ $5d$ on the other hand. The first option would go together with a (three-band) Hubbard-like description, while the second one would be more appropriately captured by a Kondo-lattice description. Moreover, it is very much possible that nontrivial competitions are present and one needs to take into account all of these degrees of freedom simultaneously. That would lead us to consider some (multiband) version of a Kondo lattice in a Hubbard model. Such a theoretical study would go beyond the present state of the art in two-dimensional (2D) calculations that do not resort to drastic approximations and it would be prudent to start by considering the respective 1D versions (to generate the rigorous results that will be needed as comparison points for more ambitious but less well controlled calculations).

\subsection{Plan of this work}

To rigorously investigate the physics of a 1D Kondo lattice in the presence of correlated band electrons is the central goal of this work. We will approach and tackle  the problem using the formalism of bosonization \cite{Giamarchi,Gogolin}. Previously, a single impurity in a Luttinger liquid was studied in detail using this technique \cite{Furusaki2005} (as well as variants of that problem \cite{Biswas2024}) and studies of a simple impurity-lattice model but in a 1D Fermi liquid were also carried out \cite{Zachar1996}. 
For the case of noninteracting hosts ($K_s=1=K_c$ in the notation to be introduced later), there is no solvable point that admits refermionization. The analysis of the bosonized model that was carried out relied on boson Green's functions and found a crossover between Kondo physics and interimpurity coherence that was sensitively dependent on the impurity concentration. For $K_c\ne1$, some parallel studies used bosonization as a starting point to set up numerical calculations (based on Monte Carlo methods) \cite{Egger1996} or focused on the half-filled case \cite{Fujimoto1997} (while we will consider other fillings).

In one dimension, one can naturally consider a generic correlated electron liquid described by a g-ology Hamiltonian, which can be efficiently described using bosonization \cite{Giamarchi,Gogolin}.
In here, we consider the combined physics %after incorporating all elements, 
of bulk interactions and spin-impurity arrays and we carry out a search for exact results that we show are possible in particular limits. Our two refermionized Toulouse points will show that, irregardless of the value of the electron-impurity magnetic exchange coupling, either of the competing physical behaviors in Doniach's balance (Kondo screening or RKKY-induced ordering) can be stabilized via the introduction of attractive or repulsive spin-channel interactions among the conduction electrons. In Doniach's language, these interactions can drive the value of $J_c$ either to zero or to infinity, respectively
%%%
\footnote{Moreover, in the limit of strong interactions, this physical picture becomes also insensitive to the concentration of impurities, provided electrons outnumber the impurities but one does not consider the limit of dilute impurities (\textit{i.e.}, $M\to\infty$ in the discussion of Toulouse point I; see Sec.\,\ref{Sec:Dilute}) for which Kondo-type screening will be found to always take precedence over impurity ordering. This is in contrast to the findings for noninteracting hosts \cite{Zachar1996}}. 
%%%

The rest of the paper is organized as follows: in Sec.\,\ref{sec:model} we present the model and describe the bosonization approach to its description; in the next section (\ref{sec:referm}) we perform a canonical transformation to simplify the model and in two subsections we identify ``Toulouse points'' at which it is possible to debosonize the model and achieve refermionized theories of noninteracting quasielectrons; the following section (\ref{sec:Green}) describes salient aspects of the physics (Green's functions, correlations, etc.); the last section (\ref{sec:conc}) presents a final discussion and the conclusions. 

\section{Model and bosonization}\label{sec:model}

We shall consider an interacting 1D spin-anisotropic Kondo lattice, consisting of a periodic array of Kondo impurities antiferromagnetically coupled to an interacting 1D electron gas. The Hamiltonian of the system is denoted by
\begin{equation}
  H= H_{e} + H_K.
\end{equation}
Here $H_{e}$ describes the dynamics of the 1D interacting electrons, which in the standard approximation of a linearized band dispersion around the Fermi level contains three terms: $H_{e}=H_0+H_e^\text{fs}+H_e^\text{bs}$, where
\begin{align} \label{eq:He}
  H_0 &= -i v_\mathrm{F} \sum_\sigma \int dx : \psi_{\sigma \mathrm{R} }^\dagger \partial_x \psi_{\sigma \mathrm{R} } - \psi_{\sigma \mathrm{L} }^\dagger \partial_x \psi_{\sigma \mathrm{L} }:\,, \\
  H_e^\text{fs} &= \sum_{\sigma\sigma',\alpha\alpha'} g_{\sigma\sigma',\alpha\alpha'} \int dx : \rho_{\sigma \alpha }\rho_{\sigma' \alpha' }:\,,\\
  H_e^\text{bs} &= g_\text{bs} \sum_{\sigma\sigma'} \delta_{\sigma',\bar{\sigma}} \int dx\, \psi_{\sigma\mathrm{L}}^{\dagger}\psi_{\sigma'\mathrm{R}}^{\dagger}\psi_{\sigma'\mathrm{L}}\psi_{\sigma\mathrm{R}}\,.\label{eq:bs}
\end{align}
$H_0$ is the normal-ordered kinetic energy of the electrons, $\psi_{\sigma R}$ and $\psi_{\sigma L}$ being annihilation operators of left- and right-moving fermions ($\alpha=-\bar\alpha\in\{\mathrm{R},\mathrm{L}\}\to\{\pm1\}$) with spin $\sigma=-\bar\sigma\in\{\uparrow,\downarrow\}\to\{\pm1\}$, and $v_\mathrm{F}$ is the Fermi velocity. $H_e^\text{fs}$ represents forward-scattering interactions, with $\rho_{\sigma\alpha}= \ :\psi^\dagger_{\sigma\alpha}\psi_{\sigma\alpha}:$ the fermionic density of each species, whereas $H_e^\text{bs}$ contains backward-scattering interactions. We have discarded umklapp scattering, assuming that the particle density is away from half filling. Notice that we are considering a general case where the interactions are not spin-rotation invariant.

The Kondo term contains two pieces, $H_K=H_\parallel+H_\nparallel$, where
\begin{align}
  H_\parallel &= J_\parallel \sum_{j}\tau_j^z \left[ \sum_\alpha \left(\rho_{\uparrow\alpha}(x_j) -  \rho_{\downarrow\alpha}(x_j)\right) \right. \label{eq:Hpara}\\
  & \left. + :e^{2i k_\mathrm{F} x_j} \left(\psi_{\uparrow\mathrm{L}}^\dagger(x_j) \psi_{ \uparrow\mathrm{R}}(x_j) - \psi_{\downarrow\mathrm{L}}^\dagger(x_j) \psi_{\downarrow\mathrm{R}}(x_j)\right)+\hc:\vphantom{\sum_\alpha}\right]\nonumber\\
  H_\nparallel &= J_\nparallel \sum_{j} : \tau_j^+ \left[\sum_\alpha \psi_{\downarrow\alpha}^\dagger(x_j) \psi_{\uparrow\alpha}(x_j)\right. \label{eq:Hperp}\\
  & +\left. e^{2i k_\mathrm{F} x_j}\left( \psi_{\downarrow \mathrm{L}}^\dagger(x_j) \psi_{\uparrow \mathrm{R}}(x_j) +  \psi_{\uparrow \mathrm{L}}^\dagger(x_j)\psi_{\downarrow \mathrm{R}}(x_j) \right)\vphantom{\sum_\alpha}\right] +\hc: \nonumber
\end{align}

Here the $\tau_{j}^a$ matrices represent the impurity spins (located at arbitrary but ordered positions $x_j$) and satisfy the canonical commutation relations $[\tau_i^a,\tau_j^b]=i\varepsilon_{abc}\tau_j^c\delta_{ij}$, with $i,j=1,\ldots,\mathcal{N_\mathrm{imp}}$, where $\mathcal{N}_\mathrm{imp}$ is the number of impurities. Eventually, we shall focus on the case of uniformly distributed impurities separated by a distance $d$, so that $x_j=jd$.

We proceed now to implement the Abelian bosonization of the model given above following a conventional procedure (cf.~Ref.~\onlinecite{shah2016,*bolech2016}). We first introduce the standard relation between chiral fermions and chiral bosons given by the bosonization identity
\begin{equation}\label{eq:chiralferm}
  \psi_{\sigma\alpha}(x) =  \frac{1}{\sqrt{2\pi a}} F_{\sigma \alpha} e^{i \alpha  \phi_{\sigma \alpha}(x)}\,
\end{equation}
where $a$ is a short distance cutoff.
The bosonic fields  $\phi_{\sigma\alpha}$ follow the usual commutation algebra
\begin{equation}\label{eq:commbos}
  \left[ \phi_{\sigma \alpha}(x) , \partial_y \phi_{\sigma' \alpha'}(y) \right] =-2  i \pi \alpha  \delta (x-y) \, \delta_{\sigma\sigma'} \delta_{\alpha\alpha'}
\end{equation}
and the Klein factors $F_{\sigma\alpha}$ satisfy the relations
\begin{align}
\{ F_{\sigma\alpha}^{\dagger},F_{\sigma'\alpha'}\} &=2\delta_{\sigma\sigma'}\delta_{\alpha\alpha'},\\
[ N_{\sigma\alpha},F_{\sigma'\alpha'}^\dagger ] &=\delta_{\sigma\sigma'}\delta_{\alpha\alpha'} F_{\sigma'\alpha'}^\dagger,\\
\{ F_{\sigma\alpha},F_{\sigma'\alpha'}\} &=0=[ F_{\sigma\alpha},\phi_{\sigma'\alpha'}],
\end{align}
where $N_{\sigma\alpha}$ are the (total) fermion-number operators
\begin{equation}
    N_{\sigma\alpha}=\int dx\,\rho_{\sigma\alpha}(x).
\end{equation} 
Following the literature \cite{Giamarchi,Gogolin}, it is customary to introduce a new set of bosonic fields through the transformation
\begin{equation}
  2 \phi_{\sigma\alpha}(x) = \Phi_c(x) + \sigma \Phi_s(x) + \bar{\alpha} \Theta_c(x) + \sigma\bar{\alpha} \Theta_s(x)
\end{equation}
which lead to dual ``spin'' (s) and ``charge'' (c) fields satisfying canonical commutation relations
\begin{equation}
    \left[ \Phi_\nu(x), \partial_y \Theta_\mu(y) \right]= 2i\pi \delta(x-y)
\end{equation}
with $\nu,\mu=c,s$. A conventional manipulation after bosonization renders the Hamiltonian of the interacting electrons into a Gaussian model for the kinetic+forward-scattering contribution, the so-called Luttinger-liquid model, $H_{LL}= H_0+H_e^\text{fs}$. This model naturally splits as $H_{LL}=H_s+H_c$, explicitly separating charge and spin modes,
\begin{equation}
   H_\nu=\frac{v_\nu}{4\pi}\int dx\,:\frac{1}{K_\nu}(\partial_x\Phi_\nu)^2+K_\nu(\partial_x\Theta_\nu)^2:,
\end{equation}
where $v_\nu$ and $K_\nu$ depend on the Fermi velocity and the forward-scattering couplings ($g_{\sigma\sigma',\alpha\alpha'}$). Physically, $v_\nu$ represents the velocity of the collective mode and $K_\nu$ is the ``Luttinger parameter'' (or stiffness parameter) that controls the decay of the correlations of that mode. In turn, the backward-scattering piece becomes
\begin{equation}
    H_e^\text{bs}=\frac{g_\text{bs}}{(2\pi a)^2} F_{\mathrm{L}\uparrow}^{\dagger}F_{\mathrm{R}\downarrow}^{\dagger}F_{\mathrm{L}\downarrow}F_{\mathrm{R}\uparrow}\int dx\, e^{2i\Phi_s}+\hc
\end{equation}

Upon introducing the bosonic representation and carrying out conventional manipulations, the terms describing the scattering of the electrons with the magnetic impurities read
\begin{widetext}
\begin{align}
  H_\parallel &= J_\parallel \sum_j \tau_j^z \left[  \frac{1}{\pi} \partial_x \Phi_{s}(x_j)\,
   + :e^{+2i k_\mathrm{F} x_j} \frac{F_{\uparrow \mathrm{L}}^\dagger F_{\uparrow \mathrm{R}}}{2\pi a} e^{+i \left(\Phi_{c}(x_j) + \Phi_{s}(x_j) \right)}
  - e^{+2i k_\mathrm{F} x_j} \frac{F_{\downarrow \mathrm{L}}^\dagger F_{\downarrow \mathrm{R}}}{2\pi a} e^{+i  \left(\Phi_{c }(x_j) - \Phi_{s}(x_j)\right)}  +\hc :\right] ,  \\
  H_\nparallel &= J_\nparallel \sum_j :\tau_j^+ \left[ \frac{F_{\downarrow \mathrm{R}}^\dagger F_{\uparrow \mathrm{R}}}{2\pi a} e^{+ i \left(\Phi_{s }(x_j) -\Theta_{s }(x_j)\right)} + \frac{F_{\downarrow \mathrm{L}}^\dagger F_{\uparrow \mathrm{L}}}{2\pi a} e^{-i   \left(\Phi_{s }(x_j) +\Theta_{s }(x_j)\right)}  \right. \nonumber \\
  & \hspace{6em}\left. +\, e^{+2i k_\mathrm{F} x_j} \frac{F_{\downarrow \mathrm{L}}^\dagger F_{\uparrow \mathrm{R}}}{2\pi a} e^{+i  \left(\Phi_{c}(x_j) - \Theta_{s}(x_j)\right)}  +e^{-2i k_\mathrm{F} x_j} \frac{F_{\downarrow \mathrm{R}}^\dagger F_{\uparrow \mathrm{L}}}{2\pi a} e^{-i \left(\Phi_{c}(x_j) + \Theta_{s}(x_j)\right)}  \right] + \hc :.
\end{align}
\end{widetext}
In this work, we will fully analyze the leading terms in both equalities given above that represent forward scattering off the impurity and give the main contribution. The remaining terms (with the oscillating exponential prefactors) describe backscattering off the impurity. In Ref.~\onlinecite{Zachar1996} it was assumed that when the electronic density is incommensurate with the lattice of impurities (i.e., $2k_Fx_j\neq n\pi$, where $n$ is the electronic density), these contributions oscillate incoherently and could be dropped. Below we shall show that, for the (range of) values of the Luttinger parameters that we will focus on, they become less relevant than the forward-scattering ones, even at commensurate fillings, and hence we will not consider them. The model one will arrive at after dropping backward-scattering terms is still highly nontrivial, since it involves a series of terms that are nonlinear in the fields and have noninteger scaling dimensions. 

\section{Solvable points\\ and Refermionization}\label{sec:referm}

In this section, we look for physically relevant solvable points. To that aim, we start by implementing a general unitary transformation of the fields in the spin sector~\cite{Emery1992,Zachar1996}
\begin{equation}\label{eq:U}
  U = \exp\left(- i \sum_j \tau_j^z \left[\lambda_\Theta \Theta_s(x_j)+\zeta \mathcal{J}_s\right] \right)\,,
\end{equation}
where $\mathcal{J}_s$ is the total spin current $\mathcal{J}_s=N_{\uparrow\mathrm{R}} - N_{\downarrow\mathrm{R}} - N_{\uparrow\mathrm{L}} + N_{\downarrow\mathrm{L}}$. This transformation produces a shift of the Luttinger Hamiltonian given by
\begin{equation}
 U^\dagger H_{LL} U = H_{LL}  - \lambda_\Theta \frac{v_s}{K_s} \sum_j \tau_j^z  \partial_x \Phi_s(x_j)
\end{equation}
and leaves the forward-scattering parallel part of the Kondo interaction unchanged (up to a constant shift).
Therefore, upon choosing $J_\parallel$ such that
\begin{equation}\label{eq:toulouse}
  \frac{J_\parallel}{\pi}-  \lambda_\Theta \frac{v_s}{K_s} = 0\,,
\end{equation}
that parallel part of the Kondo Hamiltonian is removed. The Toulouse point(s), to be introduced below, will correspond to this choice of the bare value of $J_\parallel$. The value of $\lambda_\Theta$ and the importance of the spin-current term in the transformation will become clear in the course of the refermionization procedure discussed in the following subsections. 

We still have to address $H_\nparallel$ and the other remaining terms in the Hamiltonian. To proceed, we shall assess the scaling of the different contributions to select the most relevant ones. The scaling dimensions with respect to the Luttinger-liquid fixed point (governed by $H_{LL}$) are
\begin{align}
\Delta_{\nparallel}^{\mathrm{fs}}&=\frac{1}{2}\left[K_{s}+(\lambda_{\Theta}-1)^{2}\frac{1}{K_{s}}\right]\,,\\
\Delta_{\parallel}^{\mathrm{bs}}&=\frac{1}{2}\left(K_{c}+K_{s}\right)\,,\\
\Delta_{\nparallel}^{\mathrm{bs}}&=\frac{1}{2}\left[K_{c}+(\lambda_{\Theta}-1)^{2}\frac{1}{K_{s}}\right]\,,\\
\Delta_e^\mathrm{bs}       &=2K_{s}\,. \label{eq:Deltaebs}
\end{align}
%%%
\begin{figure*}[th]
\begin{center}
\includegraphics[width=0.95\textwidth]{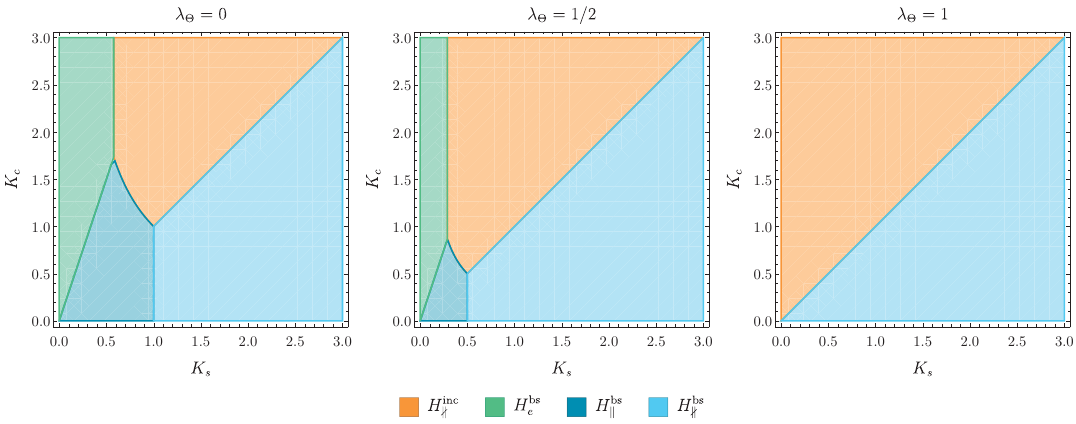}
\end{center}
\caption{\label{Fig:ScaleDims}
Regions plot in the $K_s$-$K_c$ plane indicating the term with the smallest scaling dimension. We are interested in the orange region where the incommensurate part of the spin-flip Kondo interaction, $H_\nparallel^\mathrm{inc} \to H_\nparallel^\mathrm{fs}$, is the dominant one. The different panels show how the relative importance of the different terms changes with the unitary transformation of Eq.~(\ref{eq:U}). Notice the plot for $\lambda_\Theta={3}/{2}$ would be identical to the one in the middle panel.}
\end{figure*}
%%%
Notice that these dimensions depend on $K_c$, $K_s$, and $J_\parallel$ [via $\lambda_\Theta$ through Eq.~(\ref{eq:toulouse})]. We have several possibilities to consider, but we shall be specially interested in the situation where the most relevant coupling is the one in $H_\nparallel^\text{fs}$ (see the orange region in Fig.~\ref{Fig:ScaleDims}). After the transformation, that term changes as
\begin{widetext}
\begin{multline} \label{eq:Hperpinc}
  U^\dagger H_{\nparallel}^\text{fs} U= \frac{J_\nparallel}{2\pi a} \sum_j \tau_j^+ e^{i\zeta \mathcal{J}_s} : F^\dagger_{\downarrow\mathrm{R}}F_{\uparrow\mathrm{R}}  e^{i[\Phi_s(x_j)+(\lambda_\Theta-1)\Theta_s(x_j)]}e^{i\pi\lambda_\Theta\sum_l  \tau_l^z  \sign(x_l - x_j)} e^{-2i\zeta S_z}\\
 + F^\dagger_{\downarrow\mathrm{L}}F_{\uparrow\mathrm{L}}e^{-i[\Phi_s(x_j)-(\lambda_\Theta-1)\Theta_s(x_j)]}e^{-i\pi\lambda_\Theta\sum_l  \tau_l^z  \sign(x_l - x_j)} e^{2i\zeta S_z} : + \hc\,,
\end{multline}
\end{widetext}
where $S_z= \sum_{j=1}^{\mathcal{N}_\mathrm{imp}}\tau_j^z$. 
It is tempting at this point to proceed as in Ref.~\onlinecite{Zachar1996} and choose $\lambda_\Theta=1$, which removes the field $\Theta_s$ from $H_\nparallel^\text{fs}$. However, this still leaves us with a nonquadratic Hamiltonian for $\Phi_s$. Thus we will postpone this choice to the upcoming subsections and observe that, in addition to the freedom to choose $\lambda_\Theta$, one can first tune the Luttinger parameter to a cleverly chosen value so that $H$ becomes amenable to refermionization in such a way that it is rendered bilinear. Before doing so, let us rewrite $H$ in a rescaled chiral basis,
\begin{equation}
\sqrt{2}\tilde{\phi}_{\nu\alpha}(x) = \frac{1}{\sqrt{K_\nu}}\Phi_\nu(x) + \bar{\alpha}\sqrt{K_\nu} \Theta_\nu(x)\,,
\end{equation}
satisfying analogous commutation relations as in Eq.~(\ref{eq:commbos}),
\begin{equation}
  \left[ \tilde{\phi}_{\nu \alpha}(x) , \partial_y \tilde{\phi}_{\nu' \alpha'}(y) \right] =-2  i \pi \alpha  \delta (x-y) \, \delta_{\nu\nu'} \delta_{\alpha\alpha'}
\end{equation}
for $\nu\in\{c,s\}$. This basis is more conducive to refermionization, since chiral fermions can be built as the exponential of chiral bosonic fields. We now simply have
\begin{equation}
  H_{LL}= \sum_{\nu=c,s} \frac{v_\nu}{4\pi}\int dx \sum_\alpha :  \left(\partial_x \tilde{\phi}_{\nu\alpha}(x) \right)^2 : \label{eq:finalTL}.
\end{equation}
In $H_\nparallel^\text{fs}$, on the other hand, the combinations of fields that appear in the exponents of Eq.~(\ref{eq:Hperpinc}) are rewritten as
\begin{align}
 \Phi_{s }(x_j) + (\lambda_\Theta-1) \Theta_{s }(x_j)&= \label{eq:Kcoef1}\\
    = \left(  \frac{K_s + \lambda_\Theta-1}{\sqrt{2K_s}}\right) \tilde{\phi}_{s \mathrm{L}}&(x_j) + \left(  \frac{K_s - \lambda_\Theta + 1}{\sqrt{2K_s}}\right) \tilde{\phi}_{s \mathrm{R}}(x_j)\,,\nonumber
 \end{align}
 \begin{align}
 \Phi_s(x_j) - ( \lambda_\Theta-1)\Theta_s(x_j) &= \label{eq:Kcoef2}\\
      = \left(  \frac{K_s - \lambda_\Theta + 1}{\sqrt{2K_s}}\right) \tilde{\phi}_{s \mathrm{L}}&(x_j) + \left(  \frac{K_s + \lambda_\Theta - 1}{\sqrt{2K_s}}\right) \tilde{\phi}_{s \mathrm{R}}(x_j)\,.\nonumber
\end{align}
As one can see, both exponents share the same factors and therefore we have only two expressions to consider and two parameters to determine them.

\subsection{Toulouse point I: $K_s=1/2$}

We now search for the possible values of $K_s$ and $\lambda_\Theta$ that make each factor in parentheses either 0 or $\pm1$ so that the corresponding terms in $H_\nparallel^\text{fs}$ can be refermionized by defining new spin-sector fermions as
\begin{equation}
    \tilde{\psi}_{s\alpha}=\frac{1}{\sqrt{2\pi a}}\tilde{F}_{s\alpha}e^{i\alpha\tilde{\phi}_{s\alpha}}\,.\label{eq:psitilde}
\end{equation}
Here, $\tilde{F}_{s\alpha}$ are new spin-sector Klein factors defined as
\begin{equation}
\tilde{F}_{s\alpha}=e^{ + i\frac{\zeta}{2}(N_{s\mathrm{R}}- N_{s\mathrm{L}})} F^\dagger_{\downarrow\alpha}F_{\uparrow\alpha}\,, \label{eq:Ftilde}
\end{equation}
where $N_{s\mathrm{R}}- N_{s\mathrm{L}}=\mathcal{J}_s$ and the exponential factor guarantees that Klein factors anticommute when $\zeta = \pi (2k+1), k\in \mathbb{Z}$. We seek to only have one field in each exponent, either $\tilde{\phi}_{s\mathrm{R}}$ or $\tilde{\phi}_{s\mathrm{L}}$, so that $H_\nparallel^\text{fs}$ can be written in terms of the new fermions. We find this is possible for $K_s=1/2$ and either $\lambda_\Theta={1}/{2}$ or $\lambda_\Theta={3}/{2}$. In the former case, we obtain
\begin{align}
  \Phi_{s }(x_j) + (\lambda_\Theta-1) \Theta_{s }(x_j)&=\tilde{\phi}_{s\mathrm{R}}\,,\\
  \Phi_{s }(x_j) - (\lambda_\Theta-1) \Theta_{s }(x_j)&=\tilde{\phi}_{s\mathrm{L}}
 \end{align}
 and in the latter
 \begin{align}
  \Phi_{s }(x_j) + (\lambda_\Theta-1) \Theta_{s }(x_j)&=\tilde{\phi}_{s\mathrm{L}}\,,\\
  \Phi_{s }(x_j) - (\lambda_\Theta-1) \Theta_{s }(x_j)&=\tilde{\phi}_{s\mathrm{R}}\,.
\end{align}
For $\lambda_\Theta=1/2$, the oscillating factors in Eq.~(\ref{eq:Hperpinc}) can be written as
\begin{align}
     e^{+i\frac{\pi}{2}\sum_l \tau_l^z  \sign(x_l-x_j)} &=  e^{+i\pi \sum_{l=1}^{j-1}\tau_l^z} e^{+i\frac{\pi}{2} (S^z - \tau_j^z)}(-1)^{j-1} \label{eq:extra_factor1}\,,\\
      e^{-i\frac{\pi}{2}\sum_l \tau_l^z  \sign(x_l-x_j)} &= e^{+i\pi \sum_{l=1}^{j-1} \tau_l^z} e^{-i\frac{\pi}{2} (S^z - \tau_j^z)}\,.\label{eq:extra_factor2}
\end{align}
We also rewrite the impurity spin operators $\mathbf{\tau}_j$ in terms of fermion operators $\tilde{d}_j$. This has to be done carefully, since in  order for the $\tilde{d}$ fermions to anticommute among themselves we must also incorporate a Jordan-Wigner string as follows:
\begin{subequations}\label{eq:JW}
\begin{align}
  \tilde{d}_j        &\equiv e^{-i\pi \sum_{l=1}^{j-1} \tau_l^z} \tau_j^-\,, \label{eq:JW1}\\ 
  \tilde{d}_j^\dagger &\equiv  \tau_j^+ e^{+i\pi \sum_{l=1}^{j-1} \tau_l^z}\,,\label{eq:JW2}\\ 
  \tilde{d}_j^\dagger \tilde{d}_j &= \tau_j^z+ 1/2\,.\label{eq:JW3}
\end{align}
\end{subequations}
This ensures the anticommutation relations,
\begin{align} \{ \tilde{d}_j, \tilde{d}_l^\dagger\} &= \delta_{jl}\,, && \{\tilde{d}_j, \tilde{d}_l\} = 0 = \{\tilde{d}_j^\dagger, \tilde{d}_l^\dagger\}\,. \end{align}
In addition, the first exponential factor in Eqs.~(\ref{eq:extra_factor1})~and~(\ref{eq:extra_factor2}) exactly cancels the Jordan-Wigner string from Eq.~(\ref{eq:JW}).

Now, the issue is that neither of the $\tilde{F}_{s\alpha}$ anticommute with the $\tilde{d_j}$. In order to turn them into proper fermions, we introduce the following transformation:
 \begin{align}
 d_j & = e^{ + i\frac{\zeta}{2}\left(  N_{s\mathrm{L}}- N_{s\mathrm{R}} \right)}\tilde{d}_{j}\,,\\
  F_{s\alpha}& = e^{ + i\alpha \left(\frac{\pi}{2}-2\zeta\right) N_{d}}\tilde{F}_{s\alpha}\,,
\end{align}
where $N_d$ is the (total) number operator of $d$ fermions, $N_d=S^z+\mathcal{N}_\text{imp}/2$, and $\zeta= \pi (4k+1)$ with $k\in\mathbb{Z}$ (in the following we choose $\zeta=\pi$).
Thanks to the inclusion of the spin-current term in the unitary transformation of Eq.~(\ref{eq:U}), the exponential factors involving (total) number operators cancel out and only the (constant) contribution from $\mathcal{N}_\mathrm{imp}$ remains.
Thus, as a last step, we remove those exponentials of $\mathcal{N}_\mathrm{imp}$ by performing $U(1)$ global transformations of the band fermions
\begin{equation}
  \psi_{s\alpha}\to \psi_{s\alpha} e^{-i\alpha \frac{\pi}{4}(3 \mathcal{N}_\mathrm{imp}+1)}.
\end{equation}
where $\psi_{s\alpha}$ were first obtained by replacing $\tilde{F}_{s\alpha}$ with ${F}_{s\alpha}$ in Eq.~(\ref{eq:psitilde}).
Similar transformations are needed also in the charge sector to guarantee that the impurity fermion operators anticommute with the analogous charge-sector fermions (but since the charge and spin dynamics of a Luttinger liquid are decoupled and the impurity, in the absence of backscattering, interacts only with the spin-sector band, we do not need to give explicit formulas). Moreover, these transformations do not affect $H_{LL}$, so its spin-sector part is refermionized as
\begin{equation}
H_s=-i v_\mathrm{s} \int dx : \psi_{s \mathrm{R} }^\dagger \partial_x \psi_{s \mathrm{R} } - \psi_{s\mathrm{L} }^\dagger \partial_x \psi_{s \mathrm{L} }:
\end{equation}
and analogously for the charge sector. In turn, ${H}_\nparallel^\mathrm{fs}$ becomes
\begin{equation}
  {H}_{\nparallel}^{\mathrm{fs}}  = \frac{J_{\nparallel}}{\sqrt{2\pi a}} \sum_{j} : d_{j}^{\dagger} \psi_{s\mathrm{R}}(x_{j}) (-1)^{j-1}+ d_{j}^{\dagger} \psi_{s\mathrm{L}}(x_{j})  :  + \hc \label{eq:finalModel}
\end{equation}
%%%
\noindent\!and, if the other terms from the original Hamiltonian can be neglected, the model is fully quadratic. For that, we need the scaling dimension $\Delta_\nparallel^\text{fs}$ to be the smallest one. That scenario is discussed in Fig.~\ref{Fig:ScaleDims}, where it can be seen to be the case for any value of $K_c > K_s=1/2$ (see the middle panel). This region includes the helical-Luttinger-liquid point ($K_c=1/K_s=2$), at which $H_\nparallel^\mathrm{fs}$ is the only RG-relevant interaction ($\Delta_\nparallel^\mathrm{fs}=1/2<1$) while all the others are either irrelevant ($\Delta_\parallel^\mathrm{bs},\Delta_\nparallel^\mathrm{bs}=5/4>1$) or marginal ($\Delta_e^\mathrm{bs}=1$). Instead of choosing $\lambda_\Theta={1}/{2}$ (as done above), we could have taken the value $\lambda_\Theta={3}/{2}$. In this case, even though we have to slightly modify the transformations, we obtain the same Hamiltonian and scaling dimensions. This reflects the symmetry between left and right movers from the original Hamiltonian.

It is instructive at this point to consider the case of a single impurity, for which $H_\nparallel^\mathrm{fs}$ reduces to
\begin{equation}
    {H}_{\nparallel}^{\mathrm{fs}}  = \frac{J_{\nparallel}}{\sqrt{2\pi a}}: d^{\dagger} [ \psi_{s\mathrm{R}}(0) + \psi_{s\mathrm{L}}(0) ] :  + \mathrm{H.c.}
\end{equation}
We observe a difference with respect to the single-channel Kondo model which maps onto a resonant level model with a single fermionic band mode~\cite{schlottmann1978,Kotliar1996} as the original model has only one fermion chirality after unfolding. Here, the fermion that represents the impurity couples to two bands, reflecting the existence of two chiral modes already in the original model
\footnote{Of course, for a noninteracting host, an individual chirality can be reversed and the two modes recombined into a single one coupled to the single impurity. In here, we do not take those steps because we are really interested in the multi-impurity case.}. 
From the point of view of the individual band fermions, they are coupled to the ``whole impurity,'' which is different from the two-channel Kondo model~\cite{Emery1992}, in which ``half of a fermion'' coming from the impurity decouples from the band and the models are naturally written in terms of Majorana components of the impurity degrees of freedom (the decoupled Majorana mode is seen as responsible for the zero-temperature residual entropy in that case).

Going back to the multi-impurity case, if we now use the assumption that they are equally spaced (notice that so far we could have as well had an irregular impurity array), then the Hamiltonian may be further simplified by Fourier transformation. First, we perform a gauge transformation over the $\mathrm{R}$ fields, $\psi_{s\mathrm{R}}(x)\to\psi_{s\mathrm{R}}(x)e^{-i\frac{\pi}{d}(x-d)}$, which removes the $(-1)^{j-1}$ in ${H}_{\nparallel}^{\mathrm{fs}}$ but introduces a chemical potential term for the right fermions in $H_s$:
\begin{align}
  {H}_{\nparallel}^{\mathrm{fs}} & \to \frac{J_{\nparallel}}{\sqrt{2\pi a}} \sum_{j} : d_{j}^{\dagger}\left[ \psi_{s\mathrm{R}}(x_{j}) +  \psi_{s\mathrm{L}}(x_{j}) \right] :  + \hc\,,\\
  H_s&\to H_s -\frac{\pi v_{s}}{d}\int dx:\psi_{s\mathrm{R}}^{\dagger}(x)\psi_{s\mathrm{R}}(x):\,.
\end{align}

Next, we write the Hamiltonian in $k$ space. The presence of a finite ultraviolet regulator $a$ (which is not a lattice spacing but plays a similar role) implies that for a finite system of size $L$ one would have a finite effective number of electronic sites, $\mathcal{N}_\mathrm{e} \sim L/a$, and we take $\mathcal{N}_\mathrm{e}/\mathcal{N}_\mathrm{imp} = d/a\in\mathbb{N}$.
As a consequence, if we consider periodic boundary conditions, the pseudomomenta are written as $q,k=2\pi n /L$ $(n\in\mathbb{Z})$, with $q\in [-\pi/a,\pi/a)$ for the electron ``lattice'' and $k\in [-\pi/d,\pi/d)$ for the impurity lattice. Moreover, any $q$ can be written as $q= k +K$, with $K$ being an impurity reciprocal-lattice vector given by $K = 2\pi m /d$  with $m\in [-M,M]\subset \mathbb{Z}$, where $M=\mathcal{N}_\mathrm{e}/2\mathcal{N}_\mathrm{imp}-1/2=(d/a-1)/2$. 

The Hamiltonian is diagonal in $k$ space and in the spin sector it is written as the sum of $\mathcal{N}_\mathrm{imp}$ independent one-impurity problems, $H_s+{H}_{\nparallel}^{\mathrm{fs}}=\sum_k H_k$, with
\begin{equation}
  H_k = \sum_{K,\alpha} \epsilon_{k+K,\alpha}^{s} c^\dagger_{k+K,s\alpha} c_{k+K,s\alpha} +\tilde{J}_{\nparallel}\left(d_{k}^{\dagger}c_{k+K,s\alpha}+\mathrm{H.c.}\right)\,, \label{eq:Hk}
\end{equation}
where $c_{ks\alpha}$ ($c_{ks\alpha}^\dagger$) represents the annihilation (creation) operator for $\alpha$ fermions, $\epsilon_{k+K,\alpha}^{s}= \alpha v_s (k+K) - (\alpha+1 )v_s\pi/2d$ and $\tilde{J}_{\nparallel}= J_{\nparallel}/\sqrt{2\pi ad}$. Notice the asymmetry in the dispersion relation for right and left movers, which can be traced back to the factor $(-1)^{j-1}$ in Eq.~(\ref{eq:finalModel}) and stems from the \textit{kinks} introduced by the impurity lattice. One can preserve that asymmetry while distributing it evenly between both types of movers by introducing a constant energy-dispersion shift: $\epsilon_{k+K,\alpha}^{s} \to \alpha [v_s (k+K) - v_s\pi/2d]$, which could be seen as amounting to a recentering of the Brillouin zone in a way that more naturally accounts for the impurity physics or, in terms of momentum cutoffs, to a choice that treats the emerging asymmetry between both types of movers on an even footing.

\subsection{Toulouse point II: $K_s=2$}\label{sec:Kondo}

There is another way of reaching a Toulouse point. Looking at $H_{\nparallel}$, we see that the dependence on $\Theta_{s}(x)$ is eliminated from the exponentials simply by considering $\lambda_{\Theta}=1$, just as was done by Zachar, Kivelson, and Emery for a free fermion host~\cite{Zachar1996}. Unlike in their case, our aim here is to refermionize the system's Hamiltonian. The $\Phi_{s}$ fields cannot be used to form new fermions as they do not follow ``chiral'' commutation relations as the $\phi_{\sigma\alpha}$ fields do. Nevertheless, the interactions can help us out again. Using Eqs.~(\ref{eq:Kcoef1}) and (\ref{eq:Kcoef2}) we have 
\begin{equation}\label{eq:PhiII}
\Phi_{\nu}(x)=\sqrt{\frac{K_{\nu}}{2}}\left[\tilde{\phi}_{\nu\mathrm{L}}(x)+\tilde{\phi}_{\nu\mathrm{R}}(x)\right]
\end{equation}
when expressed in terms of the rescaled fields. So, by considering $K_{s}=2$, we see that  $H_\nparallel^\mathrm{fs}$ turns into a bilinear in terms of $e^{\pm i\tilde{\phi}_{s\alpha}}$. For this value of $K_{s}$, the backscattering among electrons is irrelevant and we can safely drop it [see Eq.~(\ref{eq:Deltaebs})]. We will choose $K_c>2$, so that the backscattering off the impurity is also irrelevant and safe to ignore (cf.~Fig.~\ref{Fig:ScaleDims}). 
%%%
From the discussion of Eqs.~(\ref{eq:psitilde}) and (\ref{eq:Ftilde}) we know that
\begin{align}
\tilde{\psi}_{s\alpha}(x) & = \frac{1}{\sqrt{2\pi a}}\tilde{F}_{s\alpha}e^{i\alpha\tilde{\phi}_{s\alpha}(x)}\label{eq:psitilde2}
\end{align}
 are a good set of mutually anticommuting fermions with $\tilde{F}_{s\mathrm{\alpha}}\equiv e^{i\frac{\pi}{2}\mathcal{J}_{s}}F_{\downarrow\alpha}^{\dagger}F_{\uparrow\alpha}$. Using this definition, applying Klein-factor properties, and choosing $\zeta=0$ allows us to write $H_\nparallel^\mathrm{fs}$ as
\begin{multline}
H_{\nparallel}^\mathrm{fs}  =J_{\nparallel}\sum_{j}:\tau_{j}^{+}\left[F_{\downarrow\mathrm{L}}^{\dagger}F_{\uparrow\mathrm{L}}\tilde{\psi}_{s\mathrm{L}}^{\dagger}(x_{j})\tilde{\psi}_{s\mathrm{R}}(x_{j})e^{i\pi\sum_{l}\tau_{l}^{z}\mathrm{sgn}(x_{l}-x_{j})}\right.
 \\ \left.+F_{\downarrow\mathrm{R}}^{\dagger}F_{\uparrow\mathrm{R}}\tilde{\psi}_{s\mathrm{R}}^{\dagger}(x_{j})\tilde{\psi}_{s\mathrm{L}}(x_{j})e^{-i\pi\sum_{l}\tau_{l}^{z}\mathrm{sgn}(x_{l}-x_{j})}\right]:+\mathrm{H.c.}
\end{multline}
In order to eliminate the exponential factors involving spin-string operators, we introduce new spin operators
\begin{subequations}\label{eq:spintilde}
\begin{align} 
\tilde{\tau}_{j}^{z} & = \tau_{j}^{z}\,,\\ 
\tilde{\tau}_{j}^{+} & =\tau_{j}^{+}e^{i\pi\sum_{l\neq j}\tau_{l}^{z}}\,,\\ 
\tilde{\tau}_{j}^{-} & =\tau_{j}^{-}e^{-i\pi\sum_{l\neq j}\tau_{l}^{z}}\,,
\end{align}
\end{subequations}
which fulfill the usual SU(2) commutation relations
$\left[\tilde{\tau}_{j}^{z},\tilde{\tau}_{l}^{\pm}\right] = \pm\delta_{jl}\tilde{\tau}_{l}^{\pm}$ and 
$\left[\tilde{\tau}_{j}^{+},\tilde{\tau}_{l}^{-}\right]=2\delta_{jl}\tilde{\tau}_{l}^{z}$. After some algebra, we get
\begin{multline}
H_{\nparallel}^\mathrm{fs}  =J_{\nparallel}\sum_{j}:\tilde{\tau}_{j}^{+}(-1)^{j-1}\left[F_{\downarrow\mathrm{L}}^{\dagger}F_{\uparrow\mathrm{L}}\tilde{\psi}_{s\mathrm{L}}^{\dagger}(x_{j})\tilde{\psi}_{s\mathrm{R}}(x_{j})\right. \\ \left.+F_{\downarrow\mathrm{R}}^{\dagger}F_{\uparrow\mathrm{R}}\tilde{\psi}_{s\mathrm{R}}^{\dagger}(x_{j})\tilde{\psi}_{s\mathrm{L}}(x_{j})(-1)^{\mathcal{N}_{\mathrm{imp}}+1}\right]:+\mathrm{H.c.}
\end{multline}
The last step in the refermionization procedure involves analyzing the remaining Klein factors.
We remark that these pairs of Klein factors, $F_{\downarrow\mathrm{R}}^{\dagger}F_{\uparrow\mathrm{R}}$ and $F_{\downarrow\mathrm{L}}^{\dagger}F_{\uparrow\mathrm{L}}$, commute $[F_{\downarrow\mathrm{R}}^{\dagger}F_{\uparrow\mathrm{R}}, F_{\downarrow\mathrm{L}}^{\dagger}F_{\uparrow\mathrm{L}}]=0$ and are unitary.
$H_{\nparallel}^\mathrm{fs}$ is the only term in the Hamiltionian where they are present (and they commute with the other terms, which is expected on physical grounds due to the normal-order regularization).
As a consequence, we may choose a basis in which both operators are diagonal and their eigenvalues are pure phase factors, $F_{\downarrow\mathrm{R}}^{\dagger}F_{\uparrow\mathrm{R}} \rvert\varphi_{\mathrm{R}},\varphi_{\mathrm{L}}\rangle=e^{i\varphi_{\mathrm{R}}}\rvert\varphi_{\mathrm{R}},\varphi_{\mathrm{L}}\rangle$ and $F_{\downarrow\mathrm{L}}^{\dagger}F_{\uparrow\mathrm{L}}\rvert\varphi_{\mathrm{R}},\varphi_{\mathrm{L}}\rangle=e^{i\varphi_{\mathrm{L}}}\rvert\varphi_{\mathrm{R}},\varphi_{\mathrm{L}}\rangle$, where the states $\rvert\varphi_{\mathrm{R}},\varphi_{\mathrm{L}}\rangle$ are eigenvectors of the Klein factors and have undefined species' number, given by $\mathcal{J}_{s\alpha}=N_{\uparrow\alpha} - N_{\downarrow\alpha}$. Then,
\begin{multline}
H_{\nparallel}^\mathrm{fs}   =J_{\nparallel}\sum_{j}:(-1)^{j-1}\left[\tilde{\tau}_{j}^{+}e^{i\varphi_{\mathrm{L}}}\right.\\\left.+\tilde{\tau}_{j}^{-}e^{-i\varphi_{\mathrm{R}}}(-1)^{\mathcal{N}_{\mathrm{imp}}+1}\right]\tilde{\psi}_{s\mathrm{L}}^{\dagger}(x_{j})\tilde{\psi}_{s\mathrm{R}}(x_{j}):+\mathrm{H.c.}
\end{multline}
Finally, we simplify $H_{\nparallel}^\mathrm{fs}$ by performing the U(1) transformations $\tilde{\psi}_{s\nu}(x_{j})\to e^{-i\nu\Delta\varphi/2}\tilde{\psi}_{s\nu}(x_{j})$, with $\Delta\varphi=\left(\varphi_{\mathrm{L}}-\varphi_{\mathrm{R}}-\pi\mathcal{N}_{\mathrm{imp}}-\pi\right)/2$, and obtain
\begin{multline}
H_{\nparallel}^\mathrm{fs}=2J_{\nparallel}\sum_{j}:(-1)^{j-1}\tilde{\boldsymbol{\tau}}_{j}\cdot\hat{\mathbf{n}}_{\varphi}\left[\tilde{\psi}_{s\mathrm{L}}^{\dagger}(x_{j})\tilde{\psi}_{s\mathrm{R}}(x_{j})\right.\\\left.+\tilde{\psi}_{s\mathrm{R}}^{\dagger}(x_{j})\tilde{\psi}_{s\mathrm{L}}(x_{j})\right]:\,,\label{eq:HTp2}
\end{multline}
where $\tilde{\boldsymbol{\tau}}_{j}=\left(\tilde{\tau}_{j}^{x},\tilde{\tau}_{j}^{y},\tilde{\tau}_{j}^{z}\right)$ and $\hat{\mathbf{n}}_{\varphi}=\left(\cos\varphi,\sin\varphi,0\right)$ with $\varphi=-\left(\varphi_{\mathrm{L}}+\varphi_{\mathrm{R}}+\pi\mathcal{N}_{\mathrm{imp}}+\pi\right)/2$.

Following \textcite{Zachar1996}, we note that the $\tilde{\boldsymbol{\tau}}_{j}\cdot\hat{\mathbf{n}}_{\varphi}$ factors commute with $H$, so they can be replaced by numbers. We also note that, as a consequence of the $(-1)^{j}$ factors, the system's ground state should belong to the subspace where the impurity spins have an antiferromagnetic (AF) order along the $\hat{\mathbf{n}}_{\varphi}$ direction. Restricting to the AF subspace, we have
\begin{equation}
\left. H_{\nparallel}^{\mathrm{fs}}\right|_{\mathrm{AF}}=-J_{\nparallel}\sum_{j}{:}\left[\tilde{\psi}_{s\mathrm{L}}^{\dagger}(x_{j})\tilde{\psi}_{s\mathrm{R}}(x_{j})+\tilde{\psi}_{s\mathrm{R}}^{\dagger}(x_{j})\tilde{\psi}_{s\mathrm{L}}(x_{j})\right]{:}\,.
\end{equation}
The choice of the AF ground state, $\lvert \mathrm{AF}_\varphi \rangle$, can be better understood by considering the bosonic form, 
\begin{equation}
H_\nparallel^\mathrm{fs}= \frac{2J_{\nparallel}}{\pi a}\sum_{j}(-1)^{j-1}\tilde{\boldsymbol{\tau}}_{j}\cdot\hat{\mathbf{n}}_{\varphi}{:}\cos{\Phi}_{s}(x_{j}){:}\label{eq:Hbos}
\end{equation}
obtained by using Eq.~(\ref{eq:PhiII}) with $K_{s}=2$. If we choose the AF configuration for the spins, then the trivial classical solution,  ${\Phi}_{s}(x)=\mathrm{const.}=2n\pi$ with $n\in\mathbb{Z}$, involves no kinetic energy contribution from ${\Phi}_{s}$ and minimizes the system's energy. Any other configuration of the impurity spins involves one or more solitons to minimize $H_\nparallel^{\mathrm{fs}}$. Those nonconstant classical configurations have a positive kinetic energy contribution, and thus a positively shifted total energy as compared with the AF case.

Following the same procedure to go to momentum space as in the other Toulouse point, the spin-sector Hamiltonian in the AF subspace is rewritten as the sum of $\mathcal{N}_\mathrm{imp}$ independent one-impurity problems, $H_s+{H}_{\nparallel}^{\mathrm{fs}}=\sum_k H_k$.
When restricting to the simpler case of $M=0$ (\textit{i.e.}, densely and evenly distributed impurities) the sum on $K$ reduces to a single term and the Hamiltonian thus obtained turns into a massive Thirring model, for low-energy electrons that are now restricted to the impurity lattice. Namely,
\begin{align}
  H_s+{H}_{\nparallel}^{\mathrm{fs}} & =\sum_{\alpha=\mathrm{R},\mathrm{L}}\sum_{k}\alpha v_{s}k\,c_{ks\alpha}^{\dagger}c_{ks\alpha}-\frac{J_{\nparallel}}{d}c_{ks\alpha}^{\dagger}c_{ks\bar{\alpha}}\,,
\end{align}
where $\bar{\alpha}=R$ when $\alpha=L$ and vice versa. Notice that the coupling $J_\nparallel$ appears divided by $d$ and thus it has a well defined continuum limit.

Next, one can diagonalize the system by defining new fermionic operators $\gamma_{k\varsigma}=\mathfrak{a}_{\varsigma.k}c_{ks\mathrm{R}}-\varsigma\mathfrak{a}_{\bar{\varsigma}.k}c_{ks\mathrm{L}}$, with $\varsigma=\pm$ and $\mathfrak{a}_{+.k}=\cos\vartheta_{k}$, $\mathfrak{a}_{-.k}=\sin\vartheta_{k}$, to obtain as usual a gapped system,
\begin{equation}
H_{s}+{H}_{\nparallel}^{\mathrm{fs}}=\sum_{k,\varsigma=\pm}\varsigma E(k)\gamma_{k\varsigma}^{\dagger}\gamma_{k\varsigma}\,,
\end{equation}
with eigenvalues $ E(k)=\sqrt{\left(v_{s}k\right)^{2}+\left(J_{\nparallel}/d\right)^{2}}$ and coefficients $\cos\vartheta_{k} =\sqrt{\left(1+v_{s}k/E(k)\right)/2}$ and $\sin\vartheta_{k} =\sqrt{\left(1-v_{s}k/E(k)\right)/2}$. 

Notice that, for both Toulouse points, the charge-sector Hamiltonian is diagonal in momentum and chirality, $H_c=\sum_{\alpha=\mathrm{R},\mathrm{L}}\sum_{k}\alpha v_{c}k\,c_{kc\alpha}^{\dagger}c_{kc\alpha}$. To obtain this Hamiltonian one considers the charge part of $H_{LL}$ in terms of the rescaled bosonic fields, as shown in Eq.~(\ref{eq:finalTL}), followed Eq.~(\ref{eq:psitilde2}) to define new charge fermionic operators analogous to the spin fermionic operators and transforms into momentum space. 

\section{Green's functions\\ and spin correlations}\label{sec:Green}

In this section, we study the dynamics by focusing on impurity-operator Green's functions at the two Toulouse points. 
We will argue that the two Toulouse points correspond to different physical scenarios (RKKY-mediated ordering tendency and Kondo screening, respectively). As in other Toulouse limits, physical scales could display modified dependencies on the bare parameters of the models (like in the case of the Kondo temperature; cf.~Ref.~\onlinecite{Bolech2006,*Iucci2008}), but we nevertheless expect them to qualitatively capture the competitions and crossovers between different physical scenarios and temperature regimes
\footnote{Note that, for Kondo-type systems, the noninteracting \textit{conventionally} refermionized models serve as a first step towards capturing some of the \textit{universal} aspects of the equilibrium correlations among the impurities \cite{Ljepoja2024a,*Ljepoja2024b,*Ljepoja2024c}.}. 
With this caveat in mind, let us analyze the two solvable Toulouse points in the order we presented them above.

\subsection{Toulouse point I: $K_s =1/2$}
\label{Sec:TP1}

Since we have rewritten the impurity spin operators $\mathbf{\tau}_j$ in terms of fermion operators $\tilde{d}_j$ via Eq.~(\ref{eq:JW}), to assess the impurity correlations we compute the imaginary-time-ordered Green's functions $G_{dd}(k,\tau) = - \langle\mathcal{T}d_k^\dagger(\tau) d_k(0) \rangle$.  It is straightforward to calculate the $\tau$-space Fourier transform of $G_{dd}$ by solving the equations of motion using the Hamiltonian $H_k$ from Eq.~\eqref{eq:Hk}. One obtains
\begin{equation}\label{eq:Gdd}
  G_{dd}(k,i\omega_n)=\frac{1}{i\omega_n-\Sigma_{dd}(k,i\omega_n)},
\end{equation}
where $\omega_n$ are fermionic Matsubara frequencies and $\Sigma_{dd}$ is the impurity self-energy:
\begin{equation}\label{eq:Sdd}
  \Sigma_{dd}(k,i\omega_n)= \sum_{K\alpha}\frac{\tilde{J}_{\nparallel}^{2}}{i\omega_n -\epsilon_{k+K,\alpha}^{s}}.
\end{equation}

To access the real-space imaginary-time behavior, one looks at the full Fourier-transformed Green's function
\begin{equation}
    G_{dd}(x_{j},\tau)=\frac{d}{2\pi}\int_{-\pi/d}^{\pi/d}dk\,e^{ikx_{j}}\frac{1}{\beta}\sum_{n=-\infty}^{\infty}e^{-i\tau\omega_{n}}G_{dd}(k,i\omega_{n}).
\end{equation}
The sum over Matsubara frequencies can be carried out with the standard method of residues to formally give
\begin{equation}
    G_{dd}(x_{j},\tau)=\frac{d}{2\pi}\int_{-\pi/d}^{\pi/d}dk\,e^{ikx_{j}}G_{dd}(k,\tau)\,,
\end{equation}
where
\begin{equation}
G_{dd}(k,\tau)=-\sum_{n}\frac{e^{-\tau E_{n}(k)}}{1+e^{-\beta E_{n}(k)}}\stackrel[z=E_{n}(k)]{}{\Res}\left\{ G_{dd}(k,z)\right\}\,,
\end{equation}
with $E_{n}(k)$ being the poles of $G_{dd}(k,z)$ in the complex-$z$ plane (assuming they all have nonzero real parts and that there are no branch points). We shall call $E_{n}(k)$ the \textit{impurity energy bands}, which in the finite-$M$ case will be found to be smooth functions of $k^2$. After integrating by parts twice, one finds
\begin{multline}
G_{dd}(x_{j},\tau)=\frac{d}{2\pi}\left[\frac{e^{ikx_{j}}}{ix_{j}}G_{dd}(k,\tau)
-\frac{e^{ikx_{j}}}{(ix_{j})^{2}}\partial_k G_{dd}(k,\tau)\right]_{-\pi/d}^{\pi/d}\\
+\frac{d}{2\pi}\int_{-\pi/d}^{\pi/d}dk\,\frac{e^{ikx_{j}}}{(ix_{j})^{2}}\partial_k^2G_{dd}(k,\tau)
\end{multline}
and further integrations by parts would give rise to successive powers of $1/x_{j}$. Due to the evenness of $G_{dd}(k,\tau)$ as a function of $k$, the leading behavior is
\begin{equation}
G_{dd}(x_{j},\tau)
=\frac{d}{\pi}\frac{(-1)^j}{x_j^{2}}\partial_k G_{dd}(k=\pi/d,\tau) - \ldots
\end{equation}
which corresponds to an algebraic quasi-long-range order among the $d$ impurities mediated by the band electrons. Notice these correlations decay faster than the $(-1)^j \sqrt{\ln{|x_j|}}/x_j$ characteristic of a critical Heisenberg chain \cite{giamarchi1989}. They are, nevertheless, consistent with an RKKY scenario.

At zero temperature, the $(-1)^j/x_j$ dependence coming from the first integration by parts could survive if one of the bands was asymmetric or intersected zero in the interior of the impurity Brillouin zone (and we would find an enhanced coherence in the form of a slower decay that almost matches that of the Heisenberg model), but we will see below that this is not the case. Regardless, in either case there is a power-law decay and the qualitative character of the RKKY-like physical picture is essentially the same.

\begin{figure}[th]
\begin{center}
\includegraphics[width=0.45\textwidth]{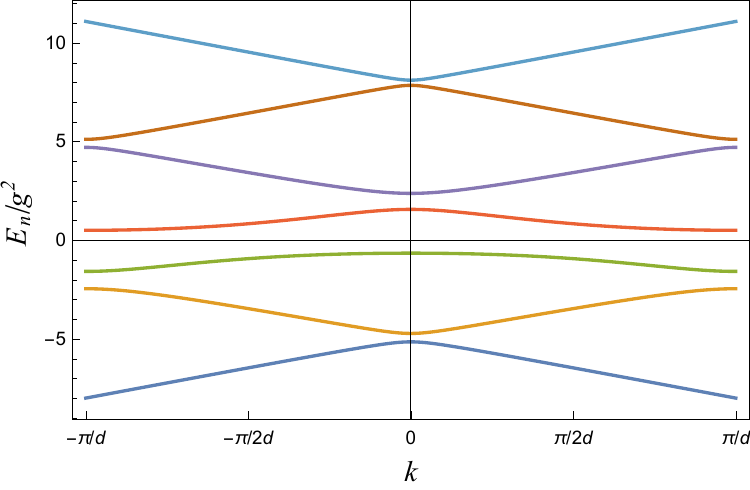}
\end{center}
\caption{\label{Fig:ImpurityBands}
Plot of the \textit{impurity energy bands} for the case of $M=1$ with $\tilde{d}=1$. The number of bands is $4M+3$, with $2M+2$ above zero and $2M+1$ below it.}
\end{figure}

We numerically find (by checking for several small values of $M$) that the impurity energy bands are real-valued even functions with no zeros. It is convenient to re\-scale momenta and frequencies in terms of $\tilde{k}=kd$ and $\tilde{z}=z/g^{2}$, with $g^{2}=\frac{d}{v_{s}}\tilde{J}_{^{\nparallel}}^{2}$. Further re\-scaling $d$ as $\tilde{d}=g^{2}d/v_{s}$, Eqs.~(\ref{eq:Gdd}) and (\ref{eq:Sdd}) yield
\begin{equation}
    g^{2}G_{dd}(\tilde{k},\tilde{z})=\frac{1}{\tilde{z}-\frac{1}{\tilde{d}}\sum_{\alpha m}\frac{1}{\tilde{z}-\frac{\alpha}{\tilde{d}}\left[\pi(2m+\frac{\alpha}{2})+\tilde{k}\right]}}\,,
\end{equation}
%\textcolor{cyan}{\begin{equation}
%    g^{2}G_{dd}(\tilde{k},\tilde{z})=\frac{1}{\tilde{z}-\frac{1}{\tilde{d}}\sum_{\alpha m}\frac{1}{\tilde{z}-\frac{\alpha}{\tilde{d}}\left[\pi(2m+\frac{\alpha+1}{2})+\tilde{k}\right]}}\,,
%\end{equation}}
where now there is a single parameter $\tilde{d}$ that affects the shape and character of the bands, $E_{n}(k)$. Rewriting the expression as a rational function and studying its denominator, we find it to be a polynomial with positive discriminant (i.e.,~whose roots are all real). The plot of the $E_{n}(k)$ shows that the bands avoid crossings and do not cross zero (see Fig.~\ref{Fig:ImpurityBands}). 

The number of bands grows with $M$ and they become flatter with increasing $\tilde{d}$ ---in particular, the (indirect) band gap around zero decreases. We could thus expect the procedure used above to sum over Matsubara frequencies to break down in the limit of large $M$ (in which $d \gg a$). Let us turn to the study of that limit.

\subsubsection{Dilute-impurities limit}
\label{Sec:Dilute}

Let us turn now to the limit when the number of sites in the ``electron lattice'' $\mathcal{N}_\mathrm{e}\to\infty$ (and $a\to 0$ keeping $L=\mathcal{N}_\mathrm{e} a$), while the number of impurity sites $\mathcal{N}_\mathrm{imp}$ stays finite. Thus $M\to\infty$ and the self-energy is given by
\begin{equation}\label{eq:Sigma_self}
  \Sigma_{dd}(k,i\omega_n) = \frac{g^2}{2}\left[\tan\frac{v_s k-i\omega_n}{2v_{s}/d}+\cot\frac{v_s k+i\omega_n}{2v_{s}/d}\right].
\end{equation}

One can read out a relevant scale in the Green's function self-energy, a frequency $\omega_{d}=v_{s}/d$ such that $\omega_{d}^{-1}$ represents the time that the spin collective excitations (the spinons) take to travel from one impurity to another. To explore other physical scales, let us consider the $k$-space Fourier transform of the Green's function. For frequencies much smaller than $\omega_{d}$ we obtain
\begin{equation}
    G_{dd}(x_{j},i\omega_n)\sim\frac{2\pi}{i\omega_nd}\left[\delta_{j,0}+\left(\frac{-\omega_n}{4g^{2}\omega_d}\right)^{x_{j}/d}+\ldots\right]\,,
\end{equation}
which, for $j\ne 0$ and focusing on the smallest-magnitude fermionic Matsubara frequency ($\omega_0 = \pi k_BT$), we can rewrite to highlight the exponential decay at long distances,
\begin{align}
    G_{dd}(x_{j},i\omega_0)\sim\frac{2\pi}{i\omega_0d}(-1)^j\,\exp\left(-\frac{x_{j}}{d}\ln\left(\frac{4g^{2}\omega_d}{\omega_0}\right)\right)\\
    \sim\frac{-2i}{k_BTd}(-1)^j\,\exp\left(-\frac{x_{j}}{d}\ln\left(T_{g}/T\right)\right)\,,
\end{align}
where we identified a characteristic temperature scale $T_{g}$, such that for $T\ll T_{g} = 4g^2 \omega_d /\pi k_B = 4\tilde{J}_{^{\nparallel}}^{2}/\pi k_B$ the expression is valid. Notice the effective correlation length increases with temperature.

A similar formula but for large frequencies gives an exponential decay,
\begin{equation}
     G_{dd}(x_{j},i\omega_n)\sim -\frac{4\pi i}{v_{s}}\, g^{2}\frac{\omega_0^2}{\omega_{n}^{2}}\left(-\frac{1}{2}\right)^{j}e^{-x_j \omega_{n}/v_{s} }.
\end{equation}
Focusing again on $\omega_0$, (or large-$\tau$ correlations),
\begin{equation}
     G_{dd}(x_{j},i\omega_0)\sim -\frac{\pi i}{v_{s}}\, 4g^{2}\left(-\frac{1}{2}\right)^{j}e^{-\frac{x_j}{(T_{d}/T)d}},
\end{equation}
where we identified a temperature scale $T_d=v_s/\pi k_B d = \omega_d/\pi k_B$ directly associated to $\omega_d$.
Notice that the correlation length now decreases with temperature for $T\gg T_{d}$. This rapid decay indicates that the spins are uncorrelated at high temperatures. 

The natural inference, after comparing the low- and high-temperature behaviors, is that there is an intermediate temperature range in which the correlation length becomes large (perhaps even briefly switching over to an algebraic quasiorder as in the finite-$M$ case), but eventually and due to the dilute limit of the impurities a local Kondo-like screening wins out and the impurities become again uncorrelated at low temperatures.

On the other hand, the large-distance equal-time correlation ($\tau=0$) at zero temperature [see Eq.~(\ref{eq:zero_time}) in the Appendix] reads
\begin{equation}
    G_{dd}(x_{j},0)\Big|_{T=0}\propto\frac{e^{-x_{j}/\xi}}{\sqrt{x_{j}}}\,,
\end{equation}
resulting in an exponential decay, with (inverse) correlation length given by
\begin{equation}
    \xi^{-1}\sim\begin{cases}\ln(6g^2)/d & \mathrm{for}\ g\gg 1\,,\\
    2\sqrt{2}\vert g\vert/d  & \mathrm{for}\ 0<|g|\ll 1\,.\end{cases}
\end{equation}
The exponential suppression of interimpurity coherence corresponds to a phase with no long-range order, not even at $T=0$ (in contrast with the algebraic order found earlier for finite $M$). This signals impurity decoupling due to Kondo-type moment screening. This interpretation is also consistent with the exponential behavior as a function of $\tau$ (for $x_j=0$) discussed in the Appendix.

\subsection{Toulouse point II: $K_s=2$}

The analysis of the correlations for Toulouse point II is qualitatively different as compared to Toulouse point I since the final Hamiltonian (obtained for the case of a dense and regular impurity array with $d/a=1$) is written fully in terms of the itinerant fermions with the understanding that the spin degrees of freedom are anti\-ferromagnetic\-ally ordered. In other words, after the transformations discussed in the previous section, the ground state is a product state of the emerging itinerant and impurity degrees of freedom (but as soon as finite temperatures are considered, the Falicov-Kimbal type of correlations will be present \cite{Falicov1969,Freericks2003,ElShawish2003}).

In order to capture the correlation physics, we start by calculating the  retarded Green's functions of the charge and spin sector fermions which are both diagonal in momentum space. For the charge part we obtain the standard free-fermion contribution 
\begin{equation}
G_{c,\alpha\gamma}^{(R)}(k,k',\omega) = \frac{\delta_{k,k'}\delta_{\alpha,\gamma}}{\omega-\alpha v_c k+i\varepsilon}
\end{equation}
while for the spin sector we have
\begin{align}
G_{s,\alpha\gamma}^{(R)}(k,k',\omega) & =\delta_{k,k'}\frac{\delta_{\alpha,\gamma}\left(\omega+\alpha v_{s}k+i\varepsilon\right)-\delta_{\alpha\bar{,\gamma}}J_{\nparallel}/d}{\left(\omega+i\varepsilon\right)^{2}-E(k)^{2}}.
\end{align}

Next, we analyze the impurity's spin-spin correlation for this Toulouse point by calculating the correlators $ \left\langle \tau_j^z(t) \tau_{ j'}^z(t')\right\rangle$ and $ \left\langle \tau_j^x(t) \tau_{ j'}^x(t')\right\rangle$. To compute the correlations using the refermionization results from Sec.~\ref{sec:Kondo} we need to transform the operators from the original Hamiltonian into their ``refermionized'' versions employing the transformations involved in the refermionization procedure. For any impurity spin operator this requires applying the transformation $U$ from Eq.~(\ref{eq:U}) with $\lambda_\Theta =1$ and $\zeta=0$, as well as the Jordan-Wigner-like map in Eq.~(\ref{eq:spintilde}). Since $[\tau_j^z, U]=0$ $\forall j$ and $\tau_j^z=\tilde{\tau}_j^z$, the impurity spin operator does not change: $\tau_j^z(t)\to\tilde{\tau}_j^z(t)$. Moreover, as we chose a ground state for the transformed impurity spin operators, they have no low-energy dynamics [i.e., $\tilde{\tau}_j^z(t)=\tilde{\tau}_j^z(0)= \tilde{\tau}_j^z$]. As a consequence, $ \left\langle \tau_j^z(t) \tau_{ j'}^z(t')\right\rangle=\left\langle \tilde{\tau}_j^z(t) \tilde{\tau}_{ j'}^z(t')\right\rangle = \left\langle \tilde{\tau}_j^z \tilde{\tau}_{ j'}^z\right\rangle $, which gives the expectation value in the system's ground-state sector given by $\lvert \mathrm{AF}_\varphi \rangle$:
\begin{equation}
  \left\langle \tau_j^z(t) \tau_{ j'}^z(t')\right\rangle \sim \frac{1}{4}\delta_{j,j'}\,.
\end{equation}
This correlator is exactly zero for $j\neq j'$, as if the spins were completely uncorrelated. But that is specific to the $\tau^z$ correlator. For the case of the $\tau^x$ or $\tau^y$ correlators, we shall see they are nonzero, albeit exponentially decreasing with the interimpurity distance.
The calculation of these correlators is more involved, since the $x$-$y$ spin operators change under the transformations presented in Eq.~(\ref{eq:spintilde}). 
In terms of the spin-ladder operators, $\tau_j^\pm(t)$, we obtain
\begin{equation}
 \left\langle \tau_j^+(t) \tau_{ j'}^{-}(t')\right\rangle = 4 \left\langle \tilde{\tau}_j^+ \tilde{\tau}_j^z \tilde{\tau}_{j'}^z \tilde{\tau}_{j'}^{-}\right\rangle \left\langle e^{i \Theta_s(x_j,t)} e^{-i \Theta_s(x_j',t')}\right\rangle
\end{equation}
by using the fact that the \textit{transformed} impurity and band parts are dynamically independent at low energies (\textit{N.B.} that the $\Theta_s$ fields arise when applying the $U$ transformation to the $\tau$'s correlator before their subsequent transformations). 
Note that the $\left\langle\tau^+\tau^+\right\rangle$ correlator is zero because its band part vanishes.
These band factors cannot be easily written in terms of $\psi_{sL}$ and $\psi_{sR}$, so we carry out the calculation in the bosonic language. 
In the dense limit, $M=0$, and, referring back to Eq.~(\ref{eq:Hbos}), the Hamiltonian can be written in terms of the bosonic fields as a free charge sector and a sine-Gordon model in the spin sector,
\begin{equation}
H=H_c+ H_{s}- \frac{J_\nparallel}{\pi a}\int dx\,:\cos\Phi_s(x):\,.
\end{equation}
In particular, the case under consideration (with $K_s=2$) corresponds to the Luther-Emery line \cite{Luther1974}, in which case the band part of the correlator displays an exponential decay in space-time \cite{Voit1998}:
\begin{equation}
  \left\langle e^{i \Theta_s(x_j,t)} e^{-i \Theta_s(x_j',t')}\right\rangle \sim \frac{\exp\left(-2 \frac{|g|}{\pi a^2}\sqrt{ \Delta x^{2}-v_{s}^{2}\Delta t^{2}}\right)}{\sqrt{\Delta x^{2}-v_{s}^{2}\Delta t^{2}}}\,,
\end{equation}
where $\Delta x=x_j-x_j'$ and $\Delta t= t-t'$. Thus we obtain the result
\begin{align}
  \left\langle \tau_j^x(t) \tau_{ j'}^x(t')\right\rangle \sim
  \frac{i}{8}(-1)^{j-j'} \frac{\exp\left(-2 \frac{|g|}{\pi a^2}\sqrt{ \Delta x^{2}-v_{s}^{2}\Delta t^{2}}\right)}{\sqrt{\Delta x^{2}-v_{s}^{2}\Delta t^{2}}}
\end{align}
signaling the absence of long range impurity order (even of an algebraic one). 

The absence of (staggered) long-range order among the impurity spins, despite the antiferromagnetic alignment of the transformed spin operators (involving their $xy$ components), might seem counterintuitive but can be understood as follows. 
The $z$-axis component of the spins is disordered and produces an exponential decay via the incoherent product of oscillatory factors coming from the unitary transformation and Jordan-Wigner strings connecting the transformed spins to the physical ones.

\section{Interpretation and Conclusions}\label{sec:conc}

The two Toulouse points identified and explored above require the electronic host of the impurities to be a strongly interacting non-Fermi liquid with a Luttinger parameter equal to either $1/2$ or $2$, respectively. Although the interactions in question are in the spin channel, let us first recall the physics for the case when the interactions are in the charge channel as that is more familiar from studying the effects of strong Coulomb interactions.

A Luttinger liquid with spin-rotation invariance (\textit{i.e.}, $K_s=1$) and with interactions in the charge channel could be arrived at as the low-energy description of a one-dimensional Hubbard model \cite{Giamarchi}. In that context, $K_c=1/2$ corresponds to the infinite-$U$ repulsive case (even smaller values of the Luttinger parameter need off-site repulsion in addition to on-site $U$). In the $U\to\infty$ case, the spin orientations of the electrons become completely random at any finite temperature and the charge dynamics is as if one had a gas of spinless fermions (called holons) due to the infinite repulsion forbidding double occupancy.
At half filling for the electrons, all sites are singly occupied in the ground state and one has a completely filled band of holons. On the other hand, $K_c=2$ corresponds to the similar limit of the large-$U$ attractive case, in which electrons with opposite spins bind strongly into pairs (called doublons). At low energies, one has a gas of doublons in which all the electron sites are either empty or doubly occupied.

To connect to the case with the interactions in the spin channel (rather than the charge channel), the two cases can be mapped into each other by exchanging the physical-sector labels of the bosons ($\phi_c \leftrightarrow \phi_s$ and thus $K_c \leftrightarrow K_s$). In terms of the original basis, this is accomplished by a change of sign of the down-spin bosons only ($\phi_\sigma \to \sigma\phi_\sigma$); or a particle-hole transformation of the down-spin fermions ($\psi_\downarrow^\dagger \to \psi_\downarrow$). The physical picture for $K_s=1/2$ is therefore a gas of neutral spinons (chargeless spin-carrying fermions) that mediate the spin exchange among impurities and give rise to the algebraic order of Toulouse point I. On the other hand, for $K_s=2$, one has bound pairs of up-spin electrons and down-spin holes (or vise versa) that, as they move in the lattice, contribute local on-site spin fluctuations and moment screening of the impurities in individual Kondo-type singlets uncorrelated from each other. This dynamics is responsible for the lack of interimpurity correlations characteristic of Toulouse point II. 

Bulk electronic interactions are thus shown to be a sensitive knob for the physics of the RKKY-Kondo transition. This would correspond, in the language of Doniach's original Kondo necklace model \cite{doniach1977}, to replacing the pseudospin chain that stands \textit{in lieu} of the noninteracting one-dimensional electron gas by an XXZ Heisenberg chain and varying the sign and strength of the bulk $z$-axis exchange coupling (such a simplified model is also amenable to bosonization at low energies and could serve as a minimal platform for numerical investigations). A hybrid analytic-numerical approach could enable systematic tuning of the interactions, interpolating between our two Toulouse limits via the Fermi-liquid bulk case; cf.\,Ref.\,\onlinecite{Egger1996}.

This conclusion could also extend to versions of the model in two or three dimensions (or when the impurity array and the electronic host have different dimensionality; \textit{e.g.}, a 1D spin necklace in a 2D host \cite{shah2003} to which interelectron interactions can be added). Those higher-dimensional cases would be relevant for a better understanding of modern materials (such as the nickelates or twisted bilayer graphene \cite{Chang2025}), but fall outside the scope of a bosonization-based analysis and should be the subject of further studies. On the other hand, there is also a growing interest in multichannel generalizations of the Kondo lattice model \cite{Wugalter2020,*Ge2022,*Ljepoja2025} and in double Kondo lattices \cite{Yang2025} (ladders in the 1D case) that could capture aspects of the multilayer physics of nickelates \cite{Zhang2025}; bosonization could be used to approach fine-tuned 1D versions of such models with the conduction band generalized from a Fermi to a Luttinger liquid.

\acknowledgments{This work was partially supported by ANPCyT,  Argentina. We acknowledge as well the hospitality of the International Center for Advanced Studies (ICAS; UNSAM, Argentina) where part of the work was carried out.}

\appendix*

\section{Fourier transform of the Matsubara function}\label{app:Greens_Matsubara}

We consider here the calculation of the Fourier transform 
\begin{equation}
G_{dd}(x_{j},\tau)=\frac{d}{2\pi}\int_{-\pi/d}^{\pi/d}dk\,\frac{1}{\beta}\sum_{n}\frac{e^{ikx_{j}}e^{-i\omega_{n}\tau}}{i\omega_{n}-\Sigma_{dd}(k,i\omega_n)},
\end{equation}
where $\Sigma_{dd}(k,i\omega_n)$ is given in Eq.~(\ref{eq:Sigma_self}), $x_j=jd$ with $j\in\mathbb{Z}$ and $\omega_n$ is the fermionic Matsubara frequency. First we eliminate the constant $d$ by changing to variables $u=kd$ and $s_{n}=\omega_{n}d/v_{s}=\omega_{n}/\omega_{d}$. Next we use the identity
\begin{equation}
    \tan\left(\frac{u-is_n}{2}\right)+\cot\left(\frac{u+is_n}{2}\right)=4i\frac{z\cosh s_n}{z^{2}-2z\sinh s_n-1}.
\end{equation}
where $z=e^{iu}$, and change variables of integration. Instead of integrating over $k$ or $u$, we go to the complex plane and integrate over $z$ counterclockwise along the unit circle; cf.\,Ref.\,\onlinecite{Alvarado2020}:  
\begin{multline}\label{eq:GreensF}
    G_{dd}(x_{j},\tau)=\frac{1}{2\pi\omega_d\beta}\sum_{n}\frac{e^{-is_{n}\omega_d\tau}}{i s_n}\oint_{C}\frac{dz}{iz}\,z^{j}\\
    \times\left[1+\frac{4zg^{2}\frac{\cosh s_{n}}{s_{n}}}{z^{2}-2z(\sinh s_{n}+2g^{2}\frac{\cosh s_{n}}{s_{n}})-1}\right]\,,
\end{multline}
where $g^{2}=J_{\nparallel}^{2}/2\pi av_{s}$ is a dimensionless coupling. The integral of the first term in the brackets is non\-vanishing only for $j=0$, so let us postpone that case. For $j>0$, we can use Cauchy's theorem to perform the integral over $z$. The poles in the denominator are located on the real axis, at
\begin{multline}
    z_{\pm}(s_n)	=\sinh s_{n}+2g^{2}\frac{\cosh s_{n}}{s_{n}}\\
    \pm\sqrt{1+\left(\sinh s_{n}+2g^{2}\frac{\cosh s_{n}}{s_{n}}\right)^{2}}\,.
\end{multline}
The product of the two roots gives $z_{+}z_{-}=1$, so there is always one inside and one outside the unit circle, and they satisfy the property $z_{+}(-s_{n})=-z_{-}(s_{n})$. Explicitly, $z_{-}$ is the one inside for $n\geqslant0$, while $z_{+}$ is inside for $n<0$. Thus we have
\begin{equation}
    \int_{C}dz\,\frac{z^{j}}{(z-z_{+})(z-z_{-})}=\begin{cases}
2\pi i\frac{z_{-}^{j}}{z_{-}-z_{+}} & n\geqslant0\,,\\
2\pi i\frac{z_{+}^{j}}{z_{+}-z_{-}} & n<0\,.
\end{cases}
\end{equation}
The summation over $n$ is difficult to carry out, but we can gain some insight by taking the zero-temperature limit. In that case, the sum is converted into an integral by means of
\begin{equation}
\frac{1}{\beta}
\sum_{n}f(\omega_{n})\to
\frac{1}{2\pi}
\int_{-\infty}^{\infty}d\omega\, f(\omega)
\end{equation}
and we obtain
\begin{multline}
    G_{dd}(x_{j},\tau)_{T=0} = \frac{1}{2\pi\omega_d}\int_{0}^{\infty}ds\,\frac{1}{is}[-z_{-}(s)]^{j} h(s)\\ 
    \times\left(\frac{e^{is\omega_d\tau}-(-1)^j e^{-is\omega_d\tau}}{2}\right)
\end{multline}
with $h(s)$ as given in Eq.~(\ref{eq:A9}).

In the $v_s\tau\ll d$ limit, we set $\tau\to 0$ and the integral vanishes for $j$ even. For $j$ odd, we use the saddle point approximation,
\begin{equation}
    \int_{0}^{\infty}ds\,h(s)e^{\lambda f(s)} \approx h(s_{0})e^{\lambda f(s_{0})}\sqrt{\frac{2\pi}{\lambda\vert f''(s_{0})\vert}}
\end{equation}
valid for large $\lambda\equiv x_j$, with $s_0$ being the location of the maximum of $f(s)$. Here we have
\begin{align}
    h(s)&=\frac{2g^{2}}{\sqrt{4g^{4}+s^{2}+4 g^{2}s\tanh s}},\label{eq:A9}\\
    f(s)&=\frac{1}{d}\ln[-z_-(s)],
\end{align}
and we obtain (for the case of $j$ odd), 
\begin{equation}\label{eq:zero_time}
    G_{dd}(x_{j},0)_{T=0}\propto\frac{e^{-{x_j}/\xi}}{\sqrt{{x_j}}}.
\end{equation}
The saddle point is determined by the solution of the equation
\begin{equation}
    \tanh s_0=\frac{1}{s_0}-\frac{s_0}{2g^2}\,,
\end{equation}
which yields
\begin{equation}
    \xi^{-1}\sim-\frac{1}{d}\ln(c/2g^2)\,,\qquad c\approx 1/3\,,
\end{equation}
for large $g$, and 
\begin{equation}
    \xi^{-1}\sim f(s_0)\sim \frac{1}{d}2|g|\sqrt{2}\,,
\end{equation}
for small $g$. 

For finite $\tau$, we carry out the integration numerically and verify a similar asymptotic exponential decay law for large $x_j$. For $j=0$ and finite $\tau$ the correlation function reads
\begin{align}\label{eq:GreensF2}
    G_{dd}(0,\tau)&=\int_{0}^{\infty}ds\,\frac{\sin s\omega_d\tau}{s\pi} \left[1-h(s)\right]\,.
\end{align}
To extract useful information from this integral --that is hard to evaluate analytically-- one can try to use the strategy of iteratively integrating by parts as we did for the momentum integration in the finite-$M$ case. In contrast to that case, here one keeps on getting zero boundary contributions at each iteration, due to the alternating parity of the two factors in the integrand, which signals a non-power-law dependence in $\tau$. This is corroborated by a numerical integration that shows an asymptotic exponential decay for large $\tau$, with a decay constant that increases monotonically with $|g|$. Such a behavior is consistent with an individual Kondo screening of each impurity spin.

\bibliography{biblio,bibliointro}

\end{document}